\documentclass[%
twocolumn,
amsmath,
amssymb,
aps,
prb,
nofootinbib,
superscriptaddress,
floatfix,
longbibliography]{revtex4-2}

\usepackage[pdftex]{graphicx}
\usepackage{hyperref}

\usepackage{bm}
\usepackage{braket}
\usepackage{amsfonts,mathrsfs}
\usepackage{mathtools}
\usepackage{xspace}
\usepackage{array}
\usepackage{multirow}
\usepackage{comment}

\hypersetup{colorlinks = true, urlcolor = blue, linkcolor = blue, citecolor = blue}

\makeatletter

\let\MYcaption\@makecaption
\makeatother

\usepackage{subcaption}
\makeatletter
\let\@makecaption\MYcaption
\makeatother

\usepackage{color}
\usepackage[normalem]{ulem} 

\begin{document}

\title{Symmetry-adapted sample-based quantum diagonalization: Application to lattice model}

\author{Kosuke Nogaki}
\email[]{kosuke.nogaki.scphys@niigata-u.ac.jp}
\affiliation{%
Department of Physics, Niigata University, Niigata 951-2181, Japan 
}
\affiliation{%
Institute for Research Administration, Niigata University, Niigata 951-2181, Japan
}
\affiliation{%
RIKEN Center for Emergent Matter Science (CEMS), Wako 351-0198, Japan
}%

\author{Steffen Backes}
\affiliation{%
RIKEN Center for Interdisciplinary Theoretical and Mathematical Sciences(iTHEMS), RIKEN, Wako 351-0198, Japan
}
\affiliation{%
RIKEN Center for Emergent Matter Science (CEMS), Wako 351-0198, Japan
}%

\author{Tomonori Shirakawa}
\affiliation{
RIKEN Center for Computational Science (R-CCS),
Kobe 650-0047, Japan
}
\affiliation{%
RIKEN Center for Quantum Computing (RQC), Wako 351-0198, Japan
}
\affiliation{%
Computational Condensed Matter Physics Laboratory, RIKEN Pioneering Research Institute (PRI), Wako 351-0198, Japan
}
\affiliation{%
RIKEN Center for Interdisciplinary Theoretical and Mathematical Sciences(iTHEMS), RIKEN, Wako 351-0198, Japan
}

\author{Seiji Yunoki}
\affiliation{%
RIKEN Center for Emergent Matter Science (CEMS), Wako 351-0198, Japan
}
\affiliation{
RIKEN Center for Computational Science (R-CCS),
Kobe 650-0047, Japan
}
\affiliation{%
RIKEN Center for Quantum Computing (RQC), Wako 351-0198, Japan
}
\affiliation{%
Computational Condensed Matter Physics Laboratory, RIKEN Pioneering Research Institute (PRI), Wako 351-0198, Japan
}
\
\author{Ryotaro Arita}
\affiliation{%
RIKEN Center for Emergent Matter Science (CEMS), Wako 351-0198, Japan
}
\affiliation{%
Department of Physics, University of Tokyo, Hongo, Tokyo 113-0033, Japan
}

\date{\today}

\begin{abstract}
We present a symmetry-adapted extension of sample-based quantum diagonalization (SQD) that rigorously embeds space-group symmetry into the many-body subspace sampled by quantum hardware.
The method is benchmarked on the two-leg ladder Hubbard model using both molecular-orbital and momentum bases. 
Energy convergence is shown to be improved in the momentum basis compared to the molecular orbital basis for both the spin-quintet ground state and the spin-singlet excited state.
We clarify the relationship between the compactness of the many-body wave function and the sparsity of the representation matrices of symmetry operations.
Furthermore, the enhancement of the superconducting correlation function due to the Coulomb interaction is demonstrated.
Our method highlights the importance of symmetry structure in random-sampling quantum simulation of correlated systems.
\end{abstract}

\maketitle

\section{Introduction}
Solving quantum many-body systems remains one of the central challenges in condensed matter physics and quantum chemistry.
The many-body wave function possesses an exponentially large number of degrees of freedom with respect to system size, and as a result, both computational cost and memory requirements quickly exceed the capabilities of classical computers.
Despite these challenges, exotic phenomena such as unconventional superconductivity~\cite{Moriya2000ap,Yanase2003pr}, magnetism~\cite{Yosida1996springer}, multipole ordering~\cite{Hayami2024jpsj}, and topological order~\cite{Wen2007book,Fradkin2013book} are widely observed in strongly correlated electron systems.
Understanding the behavior of quantum many-body systems thus remains a key objective.

Symmetry and its breaking play a fundamental role in characterizing novel quantum phases, as exemplified by Landau theory~\cite{Landau2013statistical}.
In addition to ordered phases, correlation functions can be classified according to symmetry properties.
Responses to external perturbations, transport behavior, and fluctuations are similarly governed by symmetry.
Symmetry therefore serves as a guiding principle in the study of exotic quantum phases and the functionalities of quantum materials.

The realization of pre-fault-tolerant quantum computers has stimulated growing interest in quantum computing as a promising research direction.
Quantum phase estimation is a definitive algorithm for solving quantum many-body problems, but the depth of its quantum circuits severely limits its applicability on current quantum hardware~\cite{NielsenChuang2010book,OMalley2016prx}.
As an alternative, quantum-classical hybrid diagonalization algorithms have been proposed, where the classical diagonalization step ensures noise tolerance~\cite{Kanno2023arxiv,Nakagawa2024jctc,Robledo-Moreno2024arxiv}.

In the full configuration interaction (FCI) in quantum chemistry and the exact diagonalization (ED) in condensed matter physics, the ground-state wave function $\ket{\Psi}$ is expressed as a linear combination of multiple Slater determinants to account for electron correlation effects.
The number of Slater determinants grows exponentially with the number of spin-orbitals, rendering FCI or ED calculations for large systems infeasible.
However, in many cases, $\ket{\Psi}$ exhibits significant weight concentrated on a restricted subset of Slater determinants, implying that the ground state occupies a sparse region of the full Hilbert space.
The selected configuration interaction (SCI) methods exploit this sparsity by diagonalizing the Hamiltonian in a selected subspace~\cite{Liu2016jctc,Holmes2016jctc,Tubman2016jcp}.
The quantum selected configuration interaction (QSCI) and the sample-based quantum diagonalization (SQD), a variant of SCI, use a quantum computer to sample important Slater determinants, enabling an efficient construction of the subspace.
The objective of QSCI/SQD is to approximate the ground state within a subspace where the exact ground-state wave function retains dominant amplitudes.
Following its proposal, several studies~\cite{
Barison2025qst,Liepuoniute2024arxiv,Shajan2024arxiv,Sugisaki2024,Mikkelsen2025arxiv,Nutzel2025qst,Yu2025arxiv,Yoshida2025arxiv,Danilov2025arxiv,Duriez2025arxiv, Barroca2025arxiv,Shirai2025arxiv,Ohgoe2025arxiv} have extended and investigated this approach.

QSCI/SQD was originally proposed as an alternative to the variational quantum state eigensolver (VQE).
In VQE, the energy expectation value of a variational wave function encoded on a quantum circuit is measured on a quantum computer, while classical optimization updates the variational parameters~\cite{Peruzzo2014nc,Fedorov2022mt,Cerezo2022nqi,Tilly2022pr}.
This iterative procedure seeks to minimize the expectation value of the energy.
However, noise in measurements often breaks down the variational principle, and the barren plateau problem hampers the convergence to the true ground state.
Such difficulties are ubiquitous in pre-fault-tolerant quantum devices and represent a significant limitation for VQE methods.

In this work, we propose a symmetry-adapted extension of QSCI/SQD, which enforces symmetry constraints within the selected subspace.
We apply the method to the two-leg ladder Hubbard model, a minimal model of one-dimensional unconventional superconductivity.
The symmetry properties of different basis sets in a periodic system are compared.
We demonstrate that enforcing symmetry improves the scaling of the energy expectation value as a function of the subspace dimension.
Furthermore, we evaluate superconducting correlation functions and confirm their enhancement due to the Coulomb interaction.

The rest of this paper is organized as follows. In Sec.~\ref {subsec:review}, we briefly summarize the QSCI/SQD method. 
In Sec.~\ref{subsec:symmetry-adapted}, we propose the symmetry-adapted classical postprocess scheme in the QSCI/SQD framework.
We then introduce the two-leg ladder Hubbard model in Sec.~\ref{subsec:model}.
In Sec.~\ref{subsec:two_bases}, we compare two choices of the single-particle basis from the perspective of symmetry. 
In Sec.~\ref{subsec:formulate_corre}, we explain the framework of calculating the superconducting correlation function, combined with the closed-shell scheme.
The experimental setup is summarized in Sec.~\ref{subsec:conditions}.
The acceleration of energy convergence and the behavior of correlation function are demonstrated in Sec.~\ref{subsec:energy} and Sec~\ref{subsec:correlation}, respectively.
Finally, Sec.~\ref{subsec:conclusion} summarizes this paper.

\section{Method}

\subsection{Review of Sample-based Quantum Diagonalization}
\label{subsec:review}

In this subsection, we review the methodologies referred to as QSCI~\cite{Kanno2023arxiv,Nakagawa2024jctc} and SQD~\cite{Robledo-Moreno2024arxiv}.
Given a trial ground-state wave function $\ket{\Psi_{\mathrm{trial}}}$ encoded on a quantum circuit, measurements in the computational basis yield bitstrings $\ket{\bm{x}}$.
By means of the Jordan-Wigner transformation~\cite{Jordan1928zfp}, each bitstring is mapped to a corresponding Slater determinant.
If the trial wave function shares significant support with the exact ground state, the sampled Slater determinants form an important subset for representing the ground state.
Accumulating output bitstrings enables the construction of an effective subspace $\mathcal{S} = \sum_{\bm{x}} \ket{\bm{x}}\bra{\bm{x}}$.
The projected Hamiltonian $\tilde{H} = P H P$ is then diagonalized classically to obtain an approximate ground-state energy $\tilde{E}$ and the corresponding wave function $\ket{\tilde{\Psi}}$.
In other words, $\mathcal{S}$ serves as the support of the approximate ground state.

On pre-fault-tolerant quantum hardware, noise contamination affects the sampled bitstrings, causing bit-flip errors.
Conserved quantities such as total particle number and spin number can be exploited to discard erroneous bitstrings.
However, the probability of obtaining bitstrings within the correct particle- and spin-number sector diminishes exponentially with the number of qubits.
To overcome this issue, the \textit{self-consistent recovery} technique was introduced in Ref.~\cite{Robledo-Moreno2024arxiv}.
This method systematically modifies the sampled bitstrings to enhance the fidelity of the averaged orbital occupation numbers $n_{i\sigma}$.
At each step, probabilistic bit-flips are applied based on the deviation between the erroneous bitstring and the previously averaged orbital occupations $n^{\mathrm{previous}}_{i\sigma}$.
After diagonalizing the projected Hamiltonian, new averaged occupations $n^{\mathrm{new}}_{i\sigma}$ are obtained from the approximate ground state $\ket{\tilde{\Psi}}$.
The recovery process is iterated until convergence, defined by $n^{\mathrm{new}}_{i\sigma} \approx n^{\mathrm{previous}}_{i\sigma}$.

In the sampled bitstrings, the right part represents the spin-up sector, while the left part corresponds to the spin-down sector.
A randomly sampled set of bitstrings does not guarantee that the subspace contains eigenstates of the total spin squared operator $S^2$.
To address this, the \textit{closed-shell} scheme was introduced in Ref.~\cite{Robledo-Moreno2024arxiv}.
In this approach, sampled bitstrings are separated into sets of spin-up configurations $u_a$ and spin-down configurations $u_b$, which are combined into a set of spinless configurations $\mathcal{U}$.
Spinful configurations are then reconstructed via the direct product:
\begin{equation}
  \tilde{\mathcal{S}} = \{ \tilde{u}_a \oplus \tilde{u}_b \mid \tilde{u}_a, \tilde{u}_b \in \mathcal{U} \}.
\end{equation}
The closed-shell scheme approximately restores spin-rotational symmetry within the constructed subspace.

\subsection{Symmetry Projection of Subspace}
\label{subsec:symmetry-adapted}
In many physical systems, the Hamiltonian $H$ exhibits symmetries described by a unitary operator $g$:
\begin{equation}
  H = g H g^{-1}.
\end{equation}
If $\ket{\Psi}$ is an eigenstate of $H$, then $g\ket{\Psi}$ is also an eigenstate with the same eigenvalue.
To ensure that the approximate ground state preserves this property, the effective subspace $\mathcal{S}$ should satisfy
\begin{equation}
  \mathcal{S} = g \mathcal{S} g^{-1}.
  \label{eq:symmetry_subspace}
\end{equation}
However, subspaces constructed directly from circuit measurements generally do not preserve symmetry, especially when the number of measurements is limited.

Symmetry-adapted SQD aims to restore Eq.~(\ref{eq:symmetry_subspace}) through classical postprocessing.
We begin the discussion by analyzing the action of symmetry operations on Slater determinants.
Under a symmetry operation $g$, the creation operator $c^\dagger_i$ transforms as
\begin{equation}
  g c^\dagger_i g^{-1} = \sum_j c^\dagger_j \mathcal{D}_{ji}(g),
\end{equation}
where $\mathcal{D}(g)$ is the representation matrix associated with $g$.

For a Slater determinant characterized by occupation numbers $\{n_i\}$,
\begin{align}
  \ket{n_1 n_2 \cdots n_N} &= \prod_{i=1}^{N} (c^\dagger_i)^{n_i} \ket{0} \\
  &\rightarrow \prod_{i=1}^{N} \left( \sum_j c^\dagger_j \mathcal{D}_{ji}(g) \right)^{n_i} \ket{0} \\
  &= \sum_{\{j_i\}} \prod_{i=1}^{N} (c^\dagger_{j_i} \mathcal{D}_{j_i i}(g))^{n_i} \ket{0},
  \label{eq:symmetry_projection}
\end{align}
showing that a single Slater determinant is transformed into a superposition of multiple Slater determinants.
The number of resulting determinants depends on the sparsity of $\mathcal{D}(g)$.

For each sampled bitstring $\ket{\bm{x}}$, we define an enlarged subspace
\begin{align}
  \mathcal{S}^g_{\bm{x}} = \left\{ \ket{\bm{y}} \,\middle|\, g \ket{\bm{x}} = \sum_{\bm{y}} c_{\bm{y}} \ket{\bm{y}} \right\}.
\end{align}
The symmetrized subspace is constructed as the union
\begin{align}
  \mathcal{S}^g_{\mathrm{Symmetrized}} = \bigcup_{\bm{x} \in \mathcal{S}} \mathcal{S}^g_{\bm{x}}.
\end{align}

The dimension of the symmetrized subspace, $|\mathcal{S}^g_{\mathrm{Symmetrized}}|$, exceeds that of the original subspace, $|\mathcal{S}|$.
If the representation matrix $\mathcal{D}(g)$ is sparse, the expansion ratio $|\mathcal{S}^g_{\mathrm{Symmetrized}}| / |\mathcal{S}|$ remains moderate.
From the viewpoint of computational cost, sparsity of $\mathcal{D}(g)$ is advantageous, and also suggests that the ground state retains a sparse structure.

\section{Model}

\subsection{Two-leg ladder Hubbard model}
\label{subsec:model}
\begin{figure}[tbp]
 \begin{center}
\includegraphics[keepaspectratio, scale=0.32]{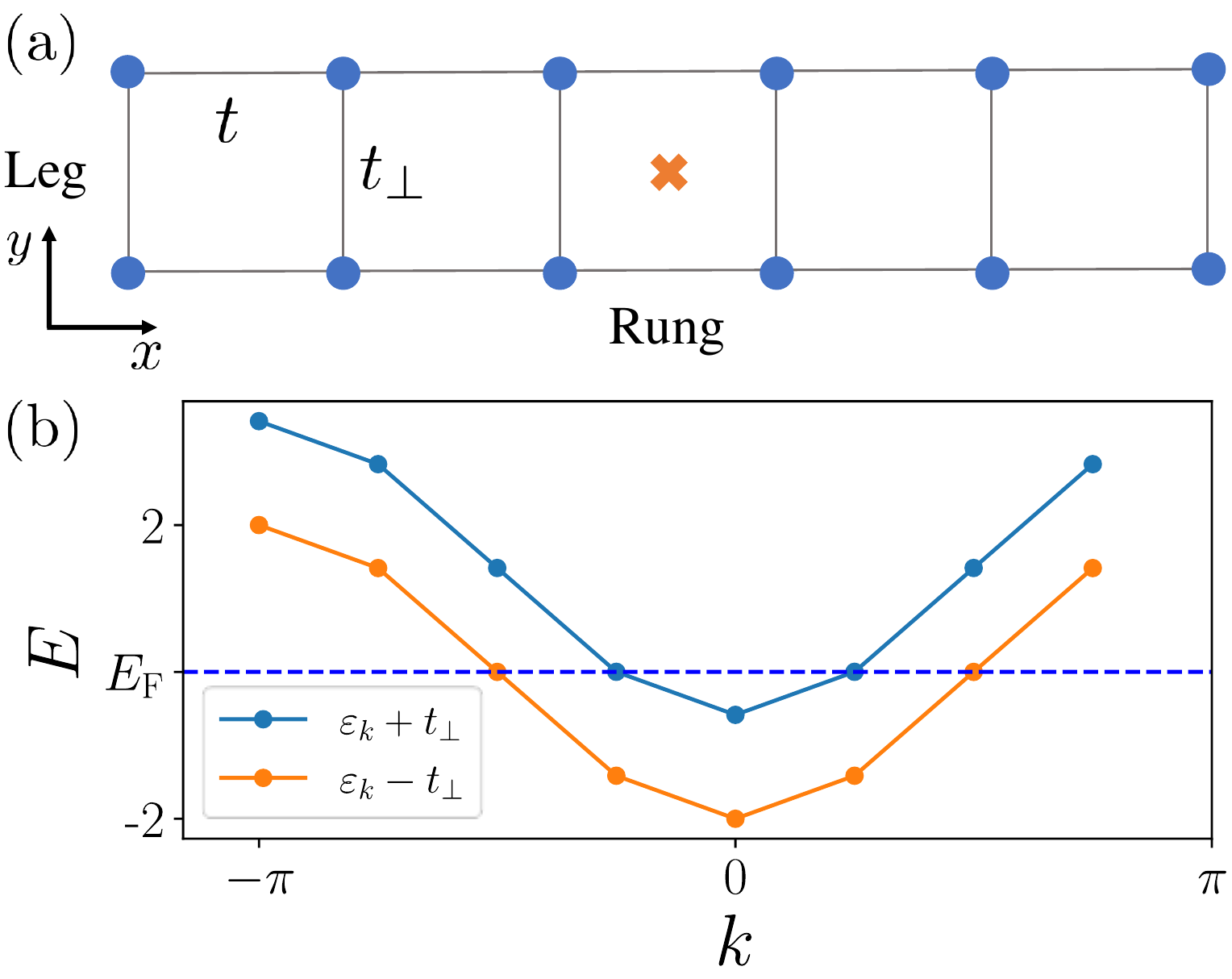}
  \end{center}
  \caption{
  (a) Lattice structure of the two-leg ladder Hubbard model.
  The rung and leg directions correspond to the $x$- and $y$-axes, respectively.
  The inversion center is indicated by the orange cross.
  (b) Band dispersion for the 8-rung case.
  The bonding and antibonding bands are shown by orange and blue lines, respectively. The Fermi energy 
  $E_{\rm F}$, corresponding to 12 electrons, is indicated by a dashed line.
  The energy difference between the highest occupied bonding level and the lowest unoccupied antibonding level is maintained below $10^{-3}$ (see Table~\ref{tab:param_2leg}).
  }
  \label{fig:two_leg_ladder}
\end{figure}

We analyze the two-leg ladder Hubbard model, a minimal model for unconventional superconductivity in strongly correlated electron systems~\cite{Dagotto1996science,Kuroki1996prb,Kimura1996prb,Kimura1997jpsj,Kimura1998jpsj}.
The lattice structure is depicted in Fig.~\ref{fig:two_leg_ladder}(a). 
This model has been extensively studied as a one-dimensional analogue of $d$-wave superconductivity in the two-dimensional square lattice.
Moreover, ladder-structured high-$T_c$ cuprates such as $\mathrm{Sr}_{n-1}\mathrm{Cu}_{n+1}\mathrm{O}_{2n}$~\cite{Hiroi1991jssc}, $\mathrm{Sr}_{14}\mathrm{Cu}_{24}\mathrm{O}_{41}$~\cite{Vuletic2006pr,Bounoua2020cp}, and $\mathrm{Sr}_x\mathrm{Ca}_{1-x}\mathrm{CuO}_2$~\cite{Rajak2023arxiv} have been discovered, highlighting the importance of this model.

The Hamiltonian is given by
\begin{align}
  H = &-t \sum_{i,s,\sigma} \left( a^\dagger_{is\sigma} a_{i+1,s\sigma} + \mathrm{H.c.} \right) \notag \\
  &- t_{\perp} \sum_{i,s} \left( a^\dagger_{is\mathrm{A}} a_{is\mathrm{B}} + \mathrm{H.c.} \right) \notag \\
  &+ U \sum_{i,\sigma} n_{i\uparrow\sigma} n_{i\downarrow\sigma},
\end{align}
where $a^\dagger_{is\sigma}$ creates an electron at rung $i$, spin $s\,(=\uparrow,\downarrow)$, and leg $\sigma\,(={\rm A},{\rm B})$, and $n_{i\sigma s} = a^\dagger_{is\sigma}a_{is\sigma}$ denotes the occupancy number operator.
The terms of Hamiltonian represent intra-leg hopping, inter-leg hopping, and on-site Coulomb interaction, respectively.
Periodic boundary conditions $c^\dagger_{i,s,\sigma} = c^\dagger_{i+N,s,\sigma}$ are imposed, where $N$ is the number of rungs.

As shown in Fig.~\ref{fig:two_leg_ladder}(b), The noninteracting band dispersion is given by
\begin{align}
  \varepsilon_{\pm}(k) = -2t\cos(k) \mp t_{\perp},
\end{align}
where the $\mp$ signs correspond to the bonding and antibonding bands, respectively.
The symmetry of the model is described by the point group $D_{2h}$.

Low-energy properties of one-dimensional many-body systems can be analyzed by bosonization and renormalization-group methods.
Using these approaches, Balents and Fisher~\cite{Balents1996prb} showed that superconducting instability, characterized by the order parameter
\begin{equation}
\mathcal{O}_i = \frac{1}{\sqrt{2}} \left( a_{i,\uparrow,\mathrm{A}} a_{i,\downarrow,\mathrm{B}} - a_{i,\downarrow,\mathrm{A}} a_{i,\uparrow,\mathrm{B}} \right),
\end{equation}
dominates when two spin channels are gapped while at least one charge channel remains gapless.
The asymptotic behavior of the superconducting correlation function is $1/r^{2K_\rho^{-1}}$, while that of the $4k_F$ charge-density-wave (CDW) correlation is $1/r^{2K_\rho}$.
Other instabilities decay exponentially.
In the weak-coupling limit, $K_\rho > 1$ has been estimated, suggesting a dominant superconducting order.

Since the presence of a finite spin gap is essential for superconductivity, the artificial gap arising from finite-size effects must be carefully managed, as pointed out in Ref.~\cite{Kuroki1996prb}.
We define the artificial gap $\Delta\varepsilon$ as the energy difference between the highest occupied and lowest unoccupied single-particle levels around the Fermi energy $E_{\mathrm{F}}$.
To mimic the thermodynamic limit, $t_\perp$ is optimized to keep $\Delta\varepsilon$ below $10^{-3}$.
The parameters used are summarized in Table~\ref{tab:param_2leg}.

\begin{table}[htbp]
  \centering
  \begin{tabular}{cccc}
    \# of rungs & \# of electrons & Filling & $t_{\perp}$ \\
    \hline
    \hline
    8 & 12 & 0.75 & 0.7076 \\
    10 & 16 & 0.80 & 1.119 \\
    12 & 20 & 0.833 & 1.366 \\
    14 & 24 & 0.857 & 0.8455 \\
    \hline
    \hline
  \end{tabular}
  \caption{Parameters for the two-leg ladder Hubbard model to ensure $\Delta\varepsilon < 10^{-3}$.
  Results for the 8-rung case are presented in Sec.~\ref{sec:result}; the 10-, 12-, and 14-rung cases are listed as references.}
  \label{tab:param_2leg}
\end{table}

\subsection{Molecular orbital basis and Momentum basis}
\label{subsec:two_bases}
In crystalline systems, the full symmetry group of the Hamiltonian is described by the space group $S$~\cite{Inui1996book,Bradley2009book}.
The space group consists of all symmetry operations combining point group and translational operations.
When the system is \textit{symmorphic}, $S$ can be expressed as the semi-direct product of the translational group $T$ and the point group $G$:
\begin{equation}
S = T \rtimes G.
\end{equation}
where $\rtimes$ denotes a semi-direct product, reflecting the fact that point group operations can map translations into other translations.
In such cases, the symmetry operations associated with translations and point group operations act independently.
This property simplifies the analysis of symmetry in the system, and allows two natural choices for the single-particle basis: one based on irreducible representations of $G$ (the molecular orbital basis) and the other based on irreducible representations of $T$ (the momentum basis).

The two-leg ladder Hubbard model exhibits symmorphic space group symmetry, and thus falls into this category.
Accordingly, in this work, we compare the molecular orbital and momentum bases from the perspectives of subspace symmetry properties and wave function sparsity.

The momentum basis is constructed by Fourier transforming the site basis:
\begin{equation}
  c^\dagger_k = \frac{1}{\sqrt{N}} \sum_i e^{ikr_i} a^\dagger_i,
\end{equation}
where $r_i$ is the position of site $i$.
Under the action of the translation operator $T_R$, which shifts all sites by $R$ to right, the momentum creation operator transforms as
\begin{equation}
  T_R c^\dagger_k T_R^{-1} = e^{-ikR} c^\dagger_k.
\end{equation}
Thus, the momentum basis forms one-dimensional irreducible representations of the translational group.
Consequently, a Slater determinant composed of momentum states transforms simply under translations:
\begin{equation}
  T_R \ket{n_{k_1} \cdots n_{k_N}} = e^{-i(\sum_i n_{k_i}k_i)R} \ket{n_{k_1} \cdots n_{k_N}}.
\end{equation}
Translational symmetry is therefore automatically preserved in subspaces randomly sampled from the momentum basis.

However, the point group symmetry is not automatically guaranteed.
In symmetry-adapted SQD, point group symmetry is recovered through a classical postprocessing step.
For the two-leg ladder Hubbard model, operations such as twofold rotation about the $y$-axis $C_{2y}$, inversion $I$, and mirror reflection $\sigma_{yz}$ act by exchanging $k \leftrightarrow -k$:
\begin{equation}
  c^\dagger_{k,\mu} \leftrightarrow c^\dagger_{-k,\mu},
\end{equation}
where $\mu$ denotes the band index (bonding or antibonding).
Other point group operations leave the momentum basis invariant.
Thus, the representation matrices of operations flipping the momentum are block-diagonal with $2\times2$ $\sigma_x$ matrices:
\begin{align}
\mathcal{D}(g) = 
\begin{pmatrix*}[c]
\sigma_x &          &        & \\
         & \sigma_x &        & \\
         &          & \ddots & \\
         &          &        & \sigma_x
\end{pmatrix*},
\label{eq:rep_mat_momentum}
\end{align}
where $\sigma_x$ is the Pauli $x$-matrix.
The representation matrix is thus highly sparse, with no diagonal elements.

To construct the molecular orbital basis, the momentum basis is projected onto irreducible representations of the point group.
The $i$-th basis function of an irreducible representation $\alpha$ is given by
\begin{equation}
  b^{\dagger,\alpha,i}_{\mu} = \mathcal{N} \frac{d_{\alpha}}{|G|} \sum_{g\in G} \mathcal{D}^{\alpha}_{ii}(g) \, g c^\dagger_{k,\mu} g^{-1},
\end{equation}
where $d_{\alpha}$ is the dimension of irreducible representation $\alpha$, $\mathcal{D}^{\alpha}(g)$ representation matrix, and $\mathcal{N}$ a normalization factor.

For the two-leg ladder Hubbard model, the molecular orbital basis reduces to
\begin{align}
  b^\dagger_{g,\mu} &= \frac{1}{\sqrt{2}} (c^\dagger_{k,\mu} + c^\dagger_{-k,\mu}), \\
  b^\dagger_{u,\mu} &= \frac{-i}{\sqrt{2}} (c^\dagger_{k,\mu} - c^\dagger_{-k,\mu}),
\end{align}
where the real-function condition is imposed.
Here, the subscripts $g$ and $u$ denote even- and odd-parity basis functions, respectively.
Since the point group $D_{2h}$ has only one-dimensional irreducible representations, each molecular orbital transforms according to a one-dimensional representation of $G$.
As a result, Slater determinants constructed from these molecular orbitals also transform as one-dimensional irreducible representations under the point group operations, and the sampled subspace based on molecular orbitals can preserve point group symmetry.

To restore the translational symmetry within the subspace, the representation matrix of the translational symmetry is needed:
\begin{align}
\mathcal{D}(T_R) = 
\begin{pmatrix*}[c]
R(k_1 R) &        &        & \\
         & R(k_2 R) &        & \\
         &        & \ddots & \\
         &        &        & R(k_{N/2} R)
\end{pmatrix*},
\label{eq:rep_mat_molecular}
\end{align}
with
\begin{equation}
R(\theta) =
\begin{pmatrix}
   \cos\theta & -\sin\theta \\
   \sin\theta & \cos\theta
\end{pmatrix}
\end{equation}
being the two-dimensional rotation matrix in the space spanned by $b^\dagger_{g,\mu}$ and $b^\dagger_{u,\mu}$.
Comparing Eqs.~(\ref{eq:rep_mat_momentum}) and (\ref{eq:rep_mat_molecular}), it is evident that the momentum basis leads to much sparser symmetry operations than the molecular orbital basis.
This greater sparsity implies that the symmetry-adapted subspace constructed from the momentum basis is more compact, resulting in improved computational efficiency for QSCI/SQD.

\subsection{Correlation function}
\label{subsec:formulate_corre}
In this subsection, we formulate the superconducting correlation function within the closed-shell framework.
Employing the closed-shell scheme, the approximate ground state is expressed as
\begin{align}
  \ket{\tilde{\Psi}} = \sum_{p,q} \tilde{\Psi}_{pq} \ket{p \oplus q},
\end{align}
where $\ket{p}$ and $\ket{q}$ are bitstrings for the spin-up and spin-down sectors, respectively.

The superconducting order parameter at site $r$ is expressed, in the molecular orbital or momentum basis, as
\begin{align}
\hat{\mathcal{O}}_r &= \sum_{i,j,\sigma,\tau} a_{i,\uparrow,\sigma} \delta_{ir} \mathcal{O}_{\sigma\tau} \delta_{rj} a_{j,\downarrow,\tau} \\
&= \sum_{\mu,\nu,i,j,\sigma,\tau} d_{i,\uparrow} (V^\top)_{\mu,i\sigma} \delta_{ir} \mathcal{O}_{\sigma\tau} \delta_{rj} V_{j\tau,\nu} d_{\nu,\downarrow} \\
&= \sum_{\mu,\nu} d_{\mu,\uparrow} \tilde{\mathcal{O}}_{\mu\nu} d_{\nu,\downarrow},
\end{align}
where $d$ denotes the annihilation operator in either the molecular orbital basis $b$ or the momentum basis $c$, and $V$ is the unitary matrix that diagonalizes the one-body Hamiltonian.
Here, $\tilde{\mathcal{O}}$ represents the transformed order parameter matrix in the computational basis.

The superconducting correlation function is then defined as
\begin{align}
  P(r) &= \braket{\tilde{\Psi}|\hat{\mathcal{O}}^\dagger_r \hat{\mathcal{O}}_0 |\tilde{\Psi}} \\
  &= \sum_{\alpha\beta\gamma\delta} \tilde{\Psi}^*_{\alpha\beta} \braket{\alpha \oplus \beta|\hat{\mathcal{O}}^\dagger_r \hat{\mathcal{O}}_0|\gamma \oplus \delta} \tilde{\Psi}_{\gamma\delta}.
\end{align}
Expanding the matrix elements explicitly, we obtain
\begin{align}
  P(r) &= \sum_{\alpha\beta\gamma\delta} \sum_{\mu\nu\xi\eta} \tilde{\Psi}^*_{\alpha\beta} (\tilde{\mathcal{O}}^\dagger_r)_{\mu\nu} (\tilde{\mathcal{O}}_0)_{\xi\eta} \tilde{\Psi}_{\gamma\delta} \notag \\
  &\quad \times \braket{\alpha \oplus \beta|d^\dagger_{\mu\downarrow} d^\dagger_{\nu\uparrow} d_{\xi\uparrow} d_{\eta\downarrow}|\gamma \oplus \delta} \\
  &= \sum_{\alpha\beta\gamma\delta} \sum_{\mu\nu\xi\eta} \tilde{\Psi}^*_{\alpha\beta} (\tilde{\mathcal{O}}^\dagger_r)_{\mu\nu} (\tilde{\mathcal{O}}_0)_{\xi\eta} \tilde{\Psi}_{\gamma\delta} \notag \\
  &\quad \times \braket{\alpha|d^\dagger_{\nu\uparrow} d_{\xi\uparrow}|\gamma}
    \braket{\beta|d^\dagger_{\mu\downarrow} d_{\eta\downarrow}|\delta}. \label{eq:cor}
\end{align}
Here, the following matrix element factorization for the four-fermion is employed,
\begin{align}
  \braket{\alpha \oplus \beta|d^\dagger_{\mu\downarrow} d^\dagger_{\nu\uparrow} d_{\xi\uparrow} d_{\eta\downarrow}|\gamma \oplus \delta}
  = \braket{\alpha|d^\dagger_{\nu\uparrow} d_{\xi\uparrow}|\gamma}
    \braket{\beta|d^\dagger_{\mu\downarrow} d_{\eta\downarrow}|\delta}.
    \label{eq:correlation_func}
\end{align}
Since $\braket{\alpha|d^\dagger_{\nu\uparrow} d_{\xi\uparrow}|\gamma}$ and $\braket{\beta|d^\dagger_{\mu\downarrow} d_{\eta\downarrow}|\delta}$ involve sparse structures, precalculating these matrix elements significantly reduces the computational cost in evaluating the full summations in Eq.~(\ref{eq:cor}) over $\alpha, \beta, \gamma, \delta$, and the orbital indices $\mu,\nu,\xi,\eta$.

\section{Result}
\label{sec:result}

\subsection{Experimental conditions}
\label{subsec:conditions}
In this study, we employed the local unitary cluster Jastrow (LUCJ) ansatz as a trial wave function $\ket{\Psi_{\mathrm{trial}}}$ on the quantum circuit~\cite{Matsuzawa2020jctc,Motta2023cs}:
\begin{align}
  \ket{\Psi_{\mathrm{LUCJ}}} = e^{\hat{T}_1 - \hat{T}_1^\dagger} e^{\hat{K}} e^{i\hat{J}} e^{-\hat{K}} \ket{\Psi_{\mathrm{RHF}}},
\end{align}
where $\ket{\Psi_{\mathrm{RHF}}}$ is the restricted Hartree-Fock (RHF) state represented in the computational basis.
Here, $\hat{T}_1$ denotes the single excitation operator from coupled-cluster singles and doubles (CCSD) theory~\cite{Asai1999prb}, and $\hat{K} = \sum_{pr,\sigma} K_{pr} c^\dagger_{p\sigma} c_{r\sigma}$ and $\hat{J} = \sum_{pr,\sigma\tau} J_{p\sigma,r\tau} n_{p\sigma} n_{r\tau}$ are one-body and density-density operators, respectively.
The amplitudes $K_{pr}$ and $J_{p\sigma,r\tau}$ are obtained through double-factorization of the CCSD two-body excitation operator $\hat{T}_2$.

The LUCJ circuit simulations are performed using the \texttt{ffsim} package~\cite{ffsim}.
The CCSD calculations during preprocessing and the iterative Davidson diagonalization during postprocessing are conducted with the quantum chemistry simulation package \texttt{PySCF}~\cite{Qiming2015jcc,Qiming2018wcms,Qiming2020jcp}.

The quantum experiments are run on IBM Quantum's \texttt{ibm\_fez}, a Heron R2 processor equipped with 156 qubits.
The characteristics of the device at the time of the investigation are presented in Table~\ref{tab:character}.
Quantum circuits are generated and transpiled using \texttt{Qiskit}~\cite{qiskit2024}.
The self-consistent recovery and subsequent diagonalization steps are carried out using \texttt{qiskit-addon-sqd}~\cite{qiskit-addon-sqd}.
The self-consistent recovery loop is repeated five times in this study.

\begin{table}[htbp]
  \centering
  \begin{tabular}{cccc}
    Property & Min. & Max. & Avg. \\
    \hline
    \hline
    $T_1$~[$\mu$s] & 22.8 & 309.9 & 141.7 \\
    $T_2$~[$\mu$s] & 5.4 & 223.1 & 90.2 \\
    Readout error & $1.71\cdot 10^{-3}$ & $1.11\cdot 10^{-1}$ & $1.44\cdot 10^{-2}$ \\
    CZ-gate error & $2.04 \cdot 10^{-3}$ & $9.92 \cdot 10^{-2}$ & $6.52 \cdot 10^{-3}$ \\
    \hline
    \hline
  \end{tabular}
  \caption{The characteristics of the IBM 156-qubit Heron R2 Processor used at the time of this study: qubit relaxation time $T_1$, qubit dephasing time $T_2$, qubit readout error and CZ-gate error. For each quantity, the minimal, maximal, and average values for the qubits used are shown in the corresponding columns.}
  \label{tab:character}
\end{table}

Since the momentum basis involves complex-valued orbitals, the diagonalization module in \texttt{PySCF} has been modified to support general two-body matrix elements~\cite{mypyscf}.
The modified version of \texttt{PySCF}~\cite{mypyscf} and a simple usage example~\cite{example} have been made publicly available.

\begin{figure}[tbp]
 \begin{center}
\includegraphics[keepaspectratio, scale=0.35]{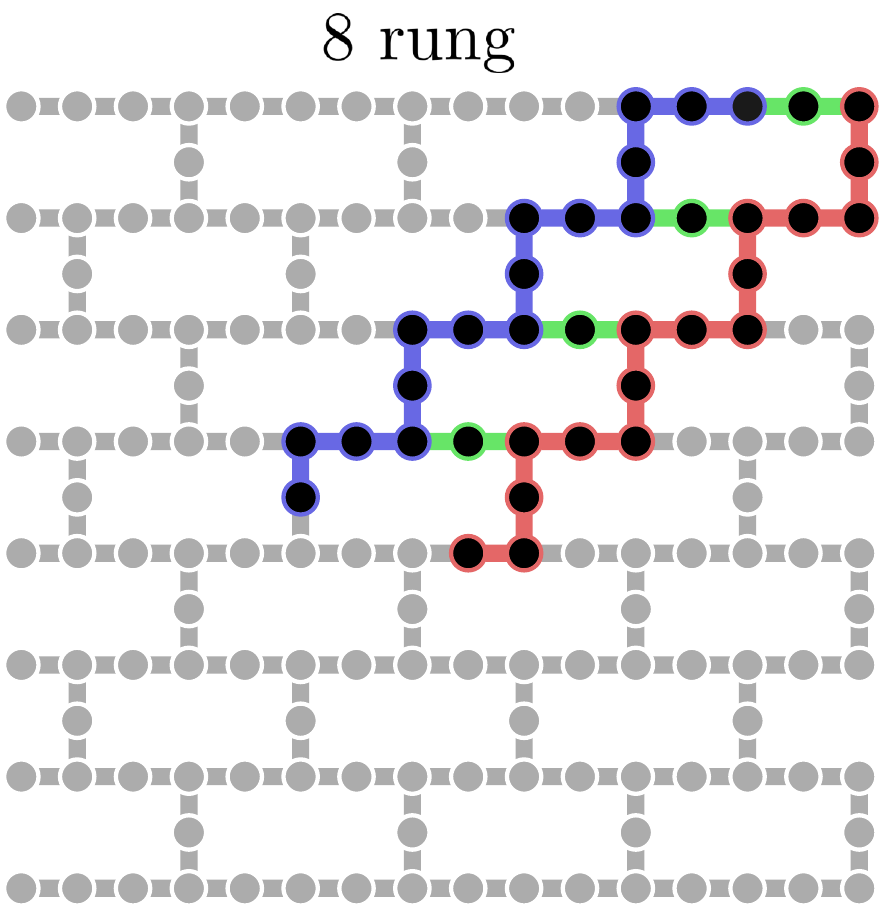}
  \end{center}
  \caption{
  Qubit layout on IBM Quantum's \texttt{ibm\_fez}.
  Blue, red, and green represent spin-up qubits, spin-down qubits, and ancilla qubits, respectively.
  8-rung system mapping is shown.
  }
  \label{fig:layout}
\end{figure}

The qubit layout used in the experiments is illustrated in Fig.~\ref{fig:layout}.
Spin-up, spin-down, and ancilla qubits are indicated by blue, red, and green, respectively.

As a reference, we also solve the model using the density-matrix renormalization group (DMRG), as implemented in \texttt{ITensor}~\cite{itensor,itensor-r0.3}. For the 8-rung model we use a truncation cutoff $\delta=10^{-7}$, resulting in a maximum bond dimension of $4340$ ($6144$) for the quintet (singlet) case.

\subsection{Convergence of energy}
\label{subsec:energy}
The DMRG calculation revealed that the 8 rungs two-leg ladder Hubbard model with our parameter choice has the spin-quintet ground state.
However, we expect that the ground state in the thermodynamic limit has the spin-singlet spin structure, since it aligns with the third law of thermodynamics.
In the following, we applied our methodology to the spin-quintet ground state and the spin-singlet excited state of the 8 rungs case.

\begin{figure}[tbp]
 \begin{center}
\includegraphics[keepaspectratio, scale=0.44]{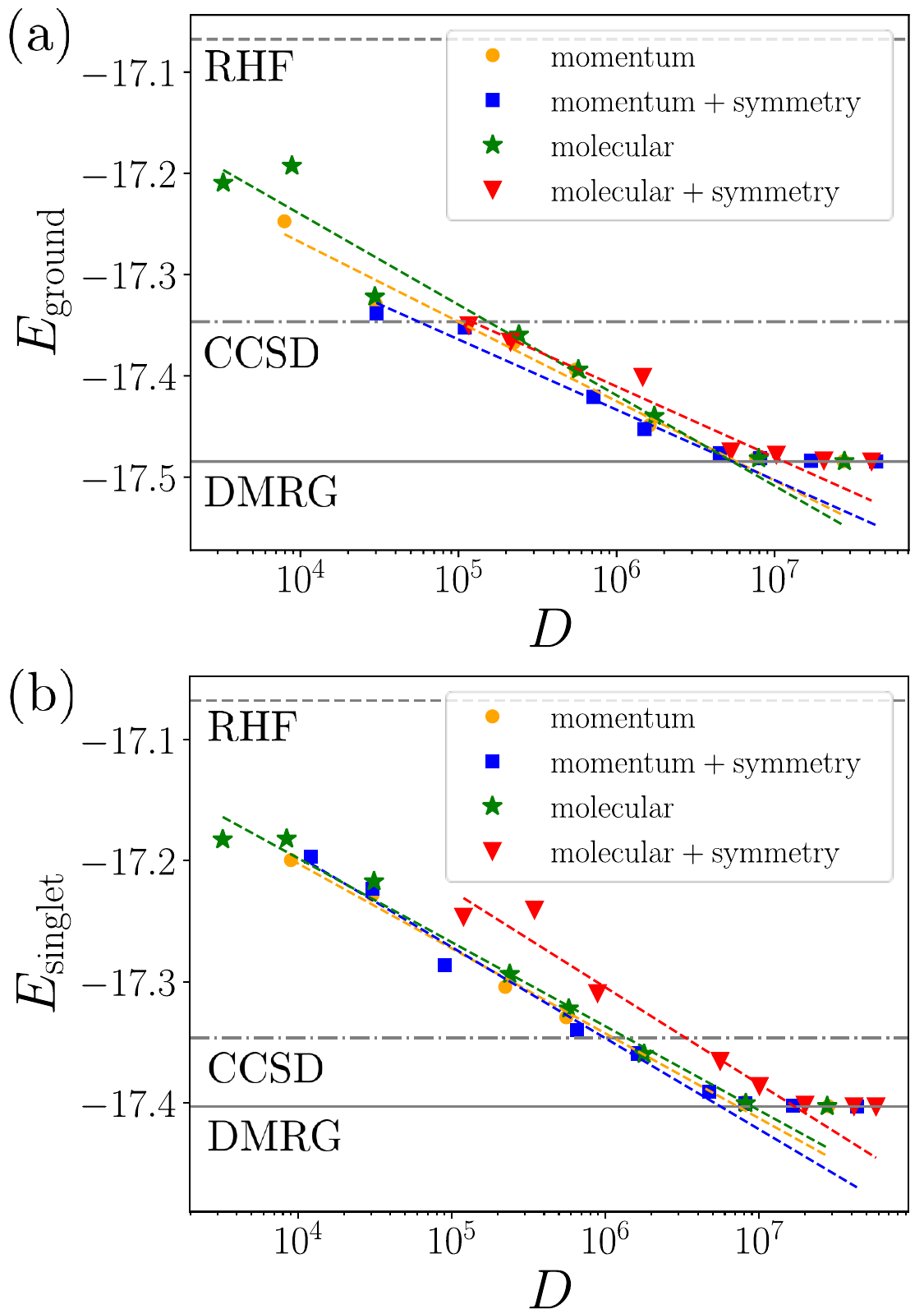}
  \end{center}
  \caption{
  Convergence of the energy expectation value as a function of subspace dimension $D$.
  Results for the molecular orbital basis and momentum basis, with and without symmetry adaptation, are shown for (a) the ground state and (b) the spin-singlet excited state. 
  The colored dashed lines are a guide to the eyes, obtained by the least-squares fitting.
  As a reference, the RHF, the CCSD, and the DMRG results are shown. 
  All results are obtained for $U/t=1$ with 12 electrons in the 8-rung ladder system.
  }
  \label{fig:energy_convergence}
\end{figure}
Figure~\ref{fig:energy_convergence} shows the convergence behavior of the energy expectation value for both the ground state of the spin quintet [Fig.~\ref{fig:energy_convergence}(a)] and the excited state of the spin singlet [Fig.~\ref{fig:energy_convergence}(b)].
Results are presented for both the molecular orbital basis and the momentum basis, each with and without applying the symmetry-adapted postprocess.

Overall, the momentum basis exhibits faster convergence compared to the molecular orbital basis, indicating a sparser structure of the wave function in the momentum basis.
Figure~\ref{fig:energy_convergence}(a) displays the results for the ground state.
In the momentum basis, enforcing symmetry using the symmetry-adapted process developed in this work improves the scaling of the energy expectation value compared to the case without symmetry enforcement.
In contrast, in the molecular orbital basis, applying the symmetry-adapted process degrades the convergence behavior.

This contrasting behavior can be understood by analyzing the structure of the representation matrices.
In the momentum basis, the representation matrix for point group operations, given in Eq.~(\ref{eq:rep_mat_momentum}), is sparse.
Consequently, in the symmetry projection formula given by Eq.~(\ref{eq:symmetry_projection}), only one term survives for each symmetry operation, effectively suppressing the growth of the symmetrized Hilbert space.
Moreover, the elements of the representation matrix are constant with respect to $k$, implying that the amplitudes of the many-body wave function are identical for a Slater determinant and its symmetry partner.
Thus, symmetry partners of important Slater determinants tend to be equally important for lowering the energy expectation value.

On the other hand, in the molecular orbital basis, the representation matrix of the translational operation given in Eq.~(\ref{eq:rep_mat_molecular}) has nonzero diagonal elements, and many terms survive in the summation of Eq.~(\ref{eq:symmetry_projection}).
Since the matrix elements depend on $k$, symmetry-related Slater determinants do not necessarily carry the same weight in the many-body wave function.
As a result, the symmetry-adapted process introduces less important configurations, leading to slower convergence.

\subsection{Superconducting correlation functions}
\label{subsec:correlation}
\begin{figure}[tbp]
 \begin{center}
\includegraphics[keepaspectratio, scale=0.3]{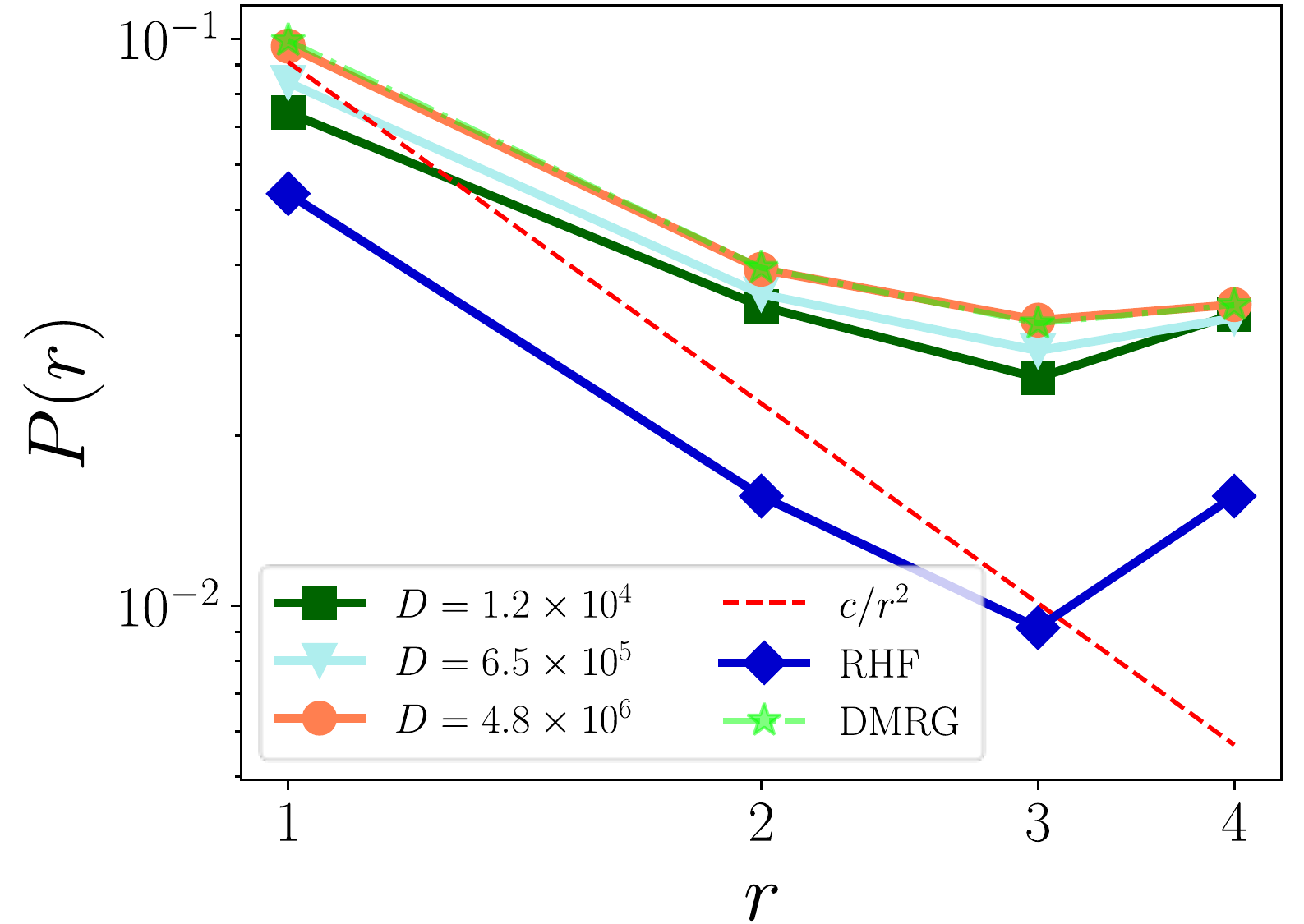}
  \end{center}
  \caption{
  The site dependence of the superconducting correlation function.
  The RHF and the interacting spin-singlet excited state results are shown. The subspace dimensions used in the symmetry-adapted QSCI/SQD in the momentum basis for the spin-singlet excited state are $D=12,100$, $D=654,481$, and $D=4,756,761$.
  The overall decay function $c/r^2$ is shown as a reference with $c=0.091$.
  All results are obtained for $U/t=1$ with 12 electrons in the 8-rung ladder system.}
  \label{fig:correlation}
\end{figure}

Next, we discuss the effect of electron correlation effect on the superconducting correlation function.
Figure~\ref{fig:correlation} presents the tail behavior of the correlation functions for the spin-singlet excited states of the various subspace dimension $D$ and the RHF state.
For comparison, the result of the DMRG calculation for the spin-singlet state is also shown.
The dashed line indicates the overall decay function $c/r^2$, where the prefactor $c$ is determined by least-squares fitting.
We find that the QSCI/SQD result for the spin-singlet state of the largest subspace dimension is in good agreement with the DMRG benchmark calculation, indicating that QSCI/SQD correctly captures the behavior of the superconducting correlation function.
Moreover, the enhanced behavior can be qualitatively captured even for relatively small subspace dimensions.

Although the system size in this study is relatively small, we observe the enhancement of the superconducting correlation function compared to the RHF state, consistent with the behavior reported in Ref.~\cite{Kuroki1996prb}.

\section{Conclusion}
\label{subsec:conclusion}
In summary, we have developed a symmetry-adapted postprocessing scheme compatible with SQD.
From a group-theoretical perspective, we have elucidated the tight connection between the compactness of the many-body wave function and the sparsity of the representation matrix of the symmetry operation.
By comparing the molecular orbital basis and the momentum basis, we have demonstrated that the momentum basis yields a more compact wave function and supports a sparser representation of the point group symmetry.

To validate our approach, we applied it to the two-leg ladder Hubbard model, a minimal model of unconventional superconductivity.
Quantum circuit simulations were performed using the LUCJ ansatz, and real-device experiments were conducted on the IBM Quantum processor.
The symmetry-adapted postprocess in the momentum basis was shown to accelerate the convergence of the energy expectation value for both the spin-quintet ground state and the spin-singlet excited state.
This acceleration can be attributed to the sparsity of the representation matrix of the symmetry operations, which enhances the effectiveness of the subspace expansion.
Finally, we demonstrated the applicability of our method by computing the superconducting correlation function within the symmetry-adapted SQD framework.
The results support the conclusion that imposing symmetry in sampled bases is crucial for improving the efficiency and accuracy of hybrid quantum-classical algorithms for strongly correlated electron systems.

\begin{acknowledgments}
K.N. was supported by JSPS KAKENHI Grant No.~JP24K22869.
T.S. was supported by JSPS KAKENHI Grant No.~JP22K03479.
S.Y. was supported by JSPS KAKENHI Grant No.~JP21H04446. 
R.A. was supported by JSPS KAKENHI Grant No.~JP25H01246.
T.S. and S.Y. were additionally grateful for funding from JST COI-NEXT (Grant No. JPMJPF2221) and the Program for Promoting Research on the Supercomputer Fugaku (Grant No. MXP1020230411) by MEXT, Japan, as well as the New Energy and Industrial Technology Development Organization (NEDO) (project No.~JPNP20017). 
T.S. and S.Y. also acknowledge support from the COE research grant in computational science from Hyogo Prefecture and Kobe City through the Foundation for Computational Science.
Furthermore, we acknowledge support from the UTokyo Quantum Initiative and the RIKEN TRIP initiative (RIKEN Quantum, Advanced General Intelligence for Science Program, Many-body Electron Systems). 
We appreciate helpful discussions with Yoshiaki Ono, Tomoki Wada, and Kazuhiro Sano.
The numerical calculations were carried out on Yukawa-21 at YITP in Kyoto University.
We appreciate helpful guidance from the \texttt{ITensor} community when writing the DMRG code.
\end{acknowledgments}

\bibliography{paper}

\end{document}